\documentclass{ws-ijmpcs}

\begin{document}

\markboth{Tam}
{The very bright and nearby GRB 130427A}

%
\catchline{}{}{}{}{}
%

\title{THE VERY BRIGHT AND NEARBY GRB 130427A: \\
The Extra Hard Spectral Component and Implications for Very High-energy Gamma-ray Observations of Gamma-ray Bursts
}

\author{Pak-Hin Thomas Tam}

\address{Institute of Astronomy and Department of Physics, National Tsing Hua University, Hsinchu 30013, Taiwan
\\
phtam@phys.nthu.edu.tw
}

\maketitle

\begin{history}
\received{31 October 2013}
\revised{20 December 2013}
\end{history}

\begin{abstract}
The extended high-energy gamma-ray ($>$100~MeV) emission occurring after the prompt gamma-ray bursts (GRBs) is usually characterized by a single power-law spectrum, which has been explained as the afterglow synchrotron radiation. We report on the Fermi Large Area Telescope (LAT) observations of the $>$100~MeV emission from the very bright and nearby GRB~130427A, up to $\sim$100~GeV. By performing time-resolved spectral fits of GRB~130427A, we found a strong evidence of an extra hard spectral component above a few GeV that exists in the extended high-energy emission of this GRB. This extra spectral component may represent the first clear evidence of the long sought-after afterglow inverse Compton emission. Prospects for observations at the very high-energy gamma-rays, i.e., above 100~GeV, are described.

\keywords{gamma-rays: observations; gamma-rays: bursts.}
\end{abstract}

\ccode{PACS numbers: 52.38.Ph, 95.85.Pw, 98.70.Rz}

\section{Large Area Telescope Observations of Gamma-ray Bursts}	
Since 2008, the Fermi Large Area Telescope (LAT) has detected more than 35 gamma-ray bursts (GRBs)\cite{lat_grb_cat} and enabled us to establish the followings:
\begin{itemlist}
 \item delayed onset of high-energy emission compared to the prompt keV-MeV emission phase,
 \item extra spectral component seen at LAT energy range during the prompt emission phase for some GRBs,
 \item the existence of the long-lasting MeV-GeV emission for many LAT GRBs.
\end{itemlist}

It is commonly believed that the long-lasting MeV-GeV emission is dominated by synchrotron radiation.\cite{zou09,kbd09,gao09,ghi10} On the other hand, the inverse Compton (IC) radiation of the forward shock electrons were expected to give rise to high-energy gamma-ray radiation as well.\cite{fan_piran_ic_review} However, the signature of IC radiation is missing: the spectrum of the extended emission above 100 MeV is usually characterized by a single power law of photon index $\sim$2 over a broad time ranges.\cite{lat_grb_cat}

\section{Properties of GRB 130427A}

The duration, $T_\mathrm{90}$, of GRB~130427A is about 138s. It has an energy fluence of 2$\times$10$^{-3}$~erg~cm$^{-2}$, putting it as the GRB with the highest fluence in GBM\cite{gcn_gbm} and Konus-Wind\cite{gcn_kw} mission lives and the most luminous GRB at z$<$0.5. It also has the highest fluence in the LAT energy range during the prompt phase.\cite{gcn_lat} Twelve $>$10~GeV photons were detected in the first 700s after the burst onset, including a 95~GeV photon that arrived at 243s after the burst onset, corresponding to an intrinsic photon energy 128~GeV at z$=$0.34, breaking previous records of the highest energy photon from GRBs~\footnote{Shortly before this oral presentation, photons with even higher energy from other GRBs were revealed using a new LAT event reconstruction scheme.\cite{pass8_grbs}}.

The 100~MeV to 100~GeV emission from GRB~130427A lasts well beyond the prompt emission period until about one day after the GRB onset. This is the longest GeV afterglow emission ever recorded for a GRB. Furthermore, the GeV photons that GRB 130427A emits contain some of the most energetic gamma-rays from GRBs. Even compared to the four brightest LAT GRBs (GRB~080916C, GRB~090510, GRB~090902B, and GRB~090926A) known so far, GRB~130427A is still peculiar in that it radiated over a hundred $>$1 GeV photons and a dozen photons at energies above 10~GeV.\cite{tam130427a}

\section{Light curve and time-resolved spectra}

\begin{figure}[pb]
\centerline{\psfig{file=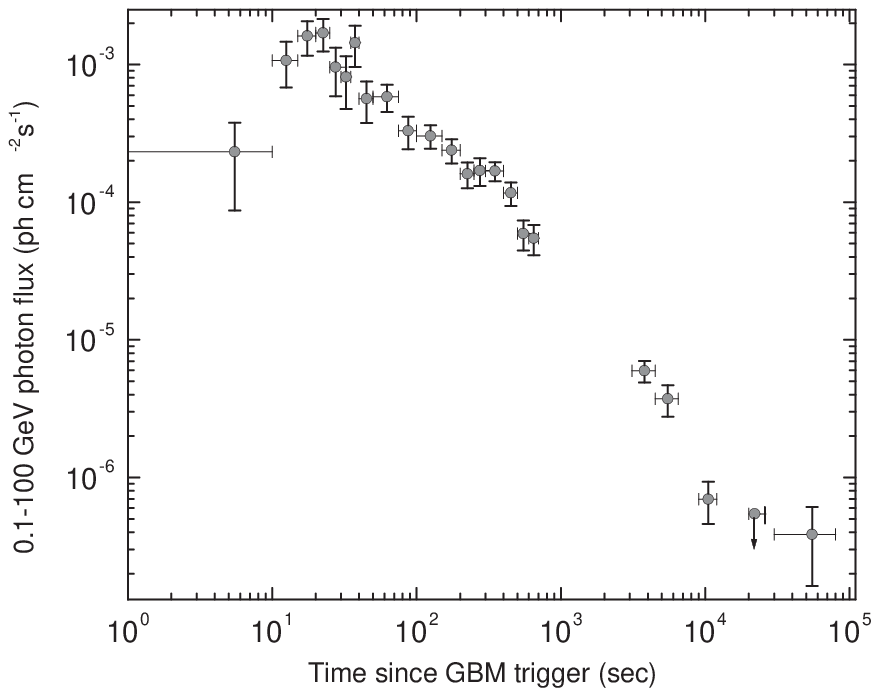,width=9.0cm}}
\vspace*{8pt}
\caption{Time evolution of the 100~MeV to 100~GeV photon flux of GRB~130427A. Reproduced from Ref.~12.\label{lc}}
\end{figure}

The 100~MeV to 100~GeV light curve derived from likelihood analysis is shown in Figure~\ref{lc}. Details of the likelihood analyzes can be found in Ref.~\refcite{fan130427a}. One can see that the peak time ($\sim$20~s), peak flux ($\sim$2$\times$10$^{-3}$~photons~cm$^{-2}$~s$^{-1}$) and the temporal decay index ($\alpha_\mathrm{L}\sim-$1.2) is not so much different from the other LAT GRBs (see Figure 14 in Ref.~\refcite{lat_grb_cat}).

\begin{figure}[pb]
\centerline{\psfig{file=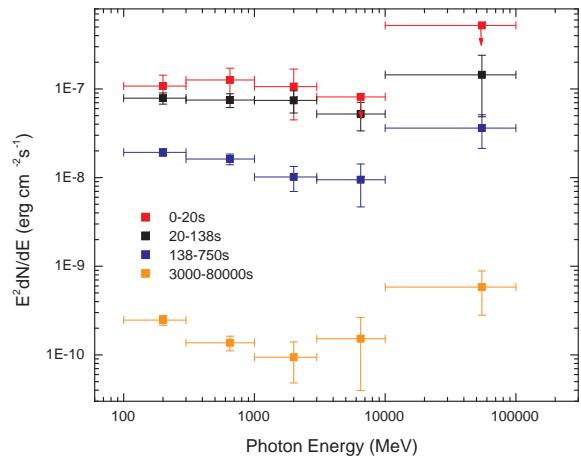,width=9.0cm}}
\vspace*{8pt}
\caption{The 100 MeV to 100 GeV spectra from the prompt emission phase to the afterglow phase. Reproduced from Ref.~11.\label{SED}}
\end{figure}

Since GRB spectra are expected to change over time, \emph{we performed the first time-resolved spectral analysis of the LAT data from GRB~130427A}.\cite{tam130427a} To extract general MeV--GeV spectral properties of GRB~130427A during the prompt and afterglow periods, only four time intervals were chosen. The results are shown in Figure~\ref{SED}. It can be clearly seen that the spectra are not well described by single power laws in some time intervals chosen here and there is a strong evidence for a spectral component below $\sim$2~GeV and another component above $\sim$2~GeV. We also tabulated the arrival times of $>$10~GeV photons.\cite{tam130427a} Fitting the data from 138~s to 80~ks with a broken power law gave an improvement over a power law at the $\sim$3 significance level. Comparison of spectral fits using single power law and broken power law are shown in Table~\ref{spec_comparism}.

\emph{It is the first time that an extra hard component above a few GeV is seen from a GRB well after the end of the prompt emission phase and up to one day after the burst.} While it is unlikely that this component is produced by the synchrotron emission, we suggest that inverse-Compton processes provide a natural alternative explanation to the photons at tens of GeV energies that arrived deep in the afterglow period.\cite{fan130427a}

\begin{table}[ph]
\tbl{Spectral properties of the GeV emission for four time intervals.}
{\begin{tabular}{ccc@{}c@{}cc} \toprule
Time since GRB onset & Power Law (PL) & & Broken Power Law (BPL) & & Improvement of BPL over PL
 \\
(sec) & $\Gamma$ & $\Gamma_\mathrm{1}$ ($E<E_\mathrm{b}$) & $\Gamma_\mathrm{2}$ ($E>E_\mathrm{b}$) & $E_\mathrm{b}$ (GeV) & ($\sigma$) \\ \colrule
    0--20 & $-$2.0$\pm$0.2 & & $...$ & & $...$ \\
    20--138 & $-$1.9$\pm$0.1 & & $...$ & & $...$ \\
    138--750 & $-$2.1$\pm$0.1 & $-$2.2$\pm$0.1 & $-$1.4$\pm$0.2 & 4.3$\pm$2.0 & 2.5 \\
    3000--80,000 & $-$2.1$\pm$0.1 & $-$2.6$\pm$0.7 & $-$1.4$\pm$0.2 & 1.1$\pm$0.9 & 2.9 \\
    138--80,000 & $-$2.1$\pm$0.1 & $-$2.3$\pm$0.2 & $-$1.4$\pm$0.1 & 2.5$\pm$1.1 & 3.5 \\ \botrule
\end{tabular} \label{spec_comparism}}
\end{table}

\section{Very High-energy Gamma-ray Observations}

\subsection{GRB~130427A}
The burst occurred when it was day time at the MAGIC and H.E.S.S. telescope sites, thus no observation was possible. It was night time for VERITAS but the full moon has prevented any observation that night.

HAWC observed it with scaler mode while the elevation of the burst was $\sim$33$^\circ$. These observing conditions caused a reduced sensitivity of the instruments and higher energy threshold. No detection was found.\cite{gcn_hawc}

\subsection{Future Prospects}
Very high-energy gamma-ray observations of other GRBs have been carried out before but no detection have been found even though some observations started just several tens of seconds after the GRB onset.\cite{hess_grb,magic_grb,veritas_grb} This can be due to several factors: (a) the relatively high redshift of many GRBs, (b) delayed observations, and/or (c) the lack of a very-high energy gamma-ray component.\cite{xue09}

We have discovered an extra hard spectral component above a few GeV from GRB~130427A that exists from $\sim$100~s up to one day after the GRB onset. This means that the afterglow spectrum of a GRB may extend to the very high-energy (VHE) gamma-ray range, i.e., $>$100~GeV, thus we have shown that the factor (c) alone is not sufficient anymore for GRBs like GRB~130427A. More GRB observations using current or future generations of the Imaging Atmosphere Cherenkov Telescopes (IACTs), such as H.E.S.S. II,\cite{hess2_grb} MAGIC II, VERITAS, and CTA, not only minutes but hours after the prompt emission phase is thus crucial to study such a hard spectral component in more details, thanks to their much larger effective collection area.

As a final note, we would like to point out that by extrapolating the LAT spectrum at late times, i.e., 3~ks to 80~ks after the burst, to very high-energy gamma-ray range, the GRB flux is still a few times stronger than the Crab nebula flux at such energies and therefore can be easily detected with, e.g., one minute of H.E.S.S. observations.

\section*{Notes added in proof}
Very recently, the Fermi LAT Collaboration also reported on the temporally extended emission.\cite{lat_130427a} Using finer time bins between 138~s and 750~s after the burst, they found that the two spectral components reported here actually doimates at different times. At late times (i.e., 3~ks to 80~ks after the burst), emission above a few GeV can be clearly seen in the spectral energy distribution during this period as well (see their Fig.~S1).

\section*{Acknowledgments}

PHT is supported by the National Science Council of the Republic of China (Taiwan) through grant NSC101-2112-M-007-022-MY3.


\end{document}